\documentclass[a4paper]{jpconf}
\begin{document}
\title{Thermodynamic Limit and Decoherence: Rigorous Results}

\author{Marco Frasca}

\address{Via Erasmo Gattamelata, 3 \\
00176 Roma (Italy)}

\ead{marcofrasca@mclink.it}

\begin{abstract}
Time evolution operator in quantum mechanics can be changed into a statistical operator by
a Wick rotation. This strict relation between statistical mechanics and quantum evolution can
reveal deep results when the thermodynamic limit is considered. These results translate in a set
of theorems proving that these effects can be effectively at work producing an emerging classical
world without recurring to any external entity that in some cases cannot be properly defined. In a
many-body system has been recently shown that Gaussian decay of the coherence is the rule with
a duration of recurrence more and more small as the number of particles increases. This effect has
been observed experimentally. More generally,
a theorem about coherence of bulk matter can be proved. All this takes us to the conclusion that
a well definite boundary for the quantum to classical world does exist and that can be drawn by
the thermodynamic limit, extending in this way the deep link between statistical mechanics and
quantum evolution to a high degree.
\end{abstract}

\section{Introduction}

The problem of measurement in quantum mechanics is one of the greatest open questions hotly
debated in physics being at the root of the way our common perceived reality is produced.
Despite a longstanding analysis, after the introduction of a concept of environmental
decoherence \cite{zeh} with an external agent that eliminates quantum correlations, the
idea of having some intrinsic effect producing the same results is still open. In the
seventies Klaus Hepp proposed the thermodynamic limit as such a cause \cite{hepp} but
some others approaches were also put forward \cite{boni,cast,ghir}. At the date the question
is not settled yet.

It is important to point out that environmental decoherence poses some serious conceptual
difficulties that motivated most of the approaches today considered. As an example, one
should consider the rather strange situation of Hyperion, a Saturn's moon, that was claimed
by Zurek \cite{zure} to become quantum after the very short period of about ten years
without environmental decoherence. This discussion is still open \cite{ball,schl}. 
 
In this paper we support the view that an instability is indeed at work in the thermodynamic
limit \cite{fra1} largely extending the initial proposal by Hepp. Indeed, both experimental
and theoretical support today exists for this view and we present here some rigorous
results about. It should be emphasized that this approach takes to completion the deep
analogy between statistical and quantum mechanics.

\section{A general theorem about bulk matter}

All the low energy phenomenology is fully described by the Hamiltonian \cite{lau}
\begin{equation}
    H = -\sum_{j=1}^{N_e}\frac{\hbar^2}{2m}\nabla^2_j
	    -\sum_{\alpha=1}^{N_i}\frac{\hbar^2}{2M_\alpha}\nabla^2_\alpha
		-\sum_{j=1}^{N_e}\sum_{\alpha=1}^{N_i}\frac{Z_\alpha e^2}{|{\bf x}_j-{\bf R}_\alpha|}
		+\sum_{j<k}\frac{e^2}{|{\bf x}_j-{\bf x}_k|}
		+\sum_{\alpha<\beta}\frac{Z_\alpha Z_\beta e^2}{|{\bf R}_\alpha-{\bf x}_\beta|}
\end{equation}
that is able to describe all the properties of matter as far as we know. We have
put $N_e$ the number of electrons, $N_i$ the number of positive ions, $e$ the electron charge,
$Z_\alpha$ the number of positive charges for each ion.

Introducing the density and neglecting for our aims the repulsion between positive ions
assumed to be pointlike and classical, the following energy functional is obtained
\begin{equation}
\label{eq:H}
    {\cal E} = T - Ze^2\int\frac{\rho(\bf x')}{|\bf x'|}d^3x'+
        \frac{e^2}{2}\int\int\frac{\rho(\bf x')\rho(\bf x'')}{|\bf x' - \bf x''|}d^3x'd^3x''
\end{equation}
being $T$ the kinetic energy given by
\begin{equation}
    T = \int d^3x_1d^2x_2d^3x_3\ldots d^3x_N\Psi^*({\bf x_1,\bf x_2,\ldots,\bf x_N})
	\left(-\frac{\hbar^2}{2m}\sum_{i=1}^N\Delta_{2i}\right)\Psi(\bf x_1,\bf x_2,\ldots,\bf x_N)
\end{equation}
being $m$ the electron mass, $\rho(\bf x)$ the electronic density computed by the Slater
determinant $\Psi$, if the Hartree-Fock approximation is invoked,
as $\int d^2x_2d^3x_3\ldots d^3x_N\Psi^*({\bf x,\bf x_2,\ldots,\bf x_N})
\Psi(\bf x,\bf x_2,\ldots,\bf x_N)$,
$Z$ is the number of positive charges and $e$ the electron charge. We assume neutrality, that is,
the number of electrons, $N=\int\rho({\bf x})d^3x$, 
is the same as the number of positive charges in the system.
One could transform the functional ${\cal E}$ into a density functional ${\cal E}[\rho(\bf x)]$,
bypassing self-consistent Hartree-Fock equations
for the Hartree-Fock approximation,
if it would be possible to obtain the kinetic energy as a functional of $\rho(\bf x)$.

We would like to accomplish such a task in the classical limit $\hbar\rightarrow 0$. We can 
reach our aim by referring to standard results in nuclear physics \cite{nt} and Bose 
condensation \cite{str} as
also given in \cite{par}: {\sl The density matrix can be developed into a $\hbar\rightarrow 0$
series, derived from the Wigner-Kirkwood series for the Green function 
for a single particle moving in the Hartree-Fock potential $V$\cite{sch}}
\begin{equation}
  G({\bf x},{\bf x'};t)=G_{TF}({\bf x},{\bf x'};t)
\left[1+\frac{\hbar^2}{12m}\left(\frac{t^2}{\hbar^2}\Delta V 
- i\frac{t^3}{\hbar^3}|\nabla V|^2\right)+\ldots\right],
\end{equation}
{\sl whose leading order is just the Thomas-Fermi approximation }
\begin{equation}
G_{TF}({\bf x},{\bf x'};t)=\left(\frac{m}{2\pi i\hbar t}\right)^{\frac{3}{2}}
\exp\left[\frac{im({\bf x}-{\bf x'})^2}{2\hbar t}
-\frac{it}{\hbar}V\left(\frac{{\bf x}+{\bf x'}}{2}\right)\right]
\end{equation}
{\sl giving the density of kinetic energy as}
\begin{equation}
    \tau_{TF} = \frac{3}{10}\frac{\hbar^2}{m}(3\pi^2)^{\frac{2}{3}}\rho({\bf x})^{\frac{5}{3}}.
\end{equation}
So, we can conclude that the Thomas-Fermi approximation can represent classical objects being the
leading order of a semiclassical approximation.

For our aims this is not enough as we are interested in the limit of a large number of particles 
in the energy functional we started with. Indeed, there is another theorem due to Lieb and 
Simon\cite{ls1,ls2} that gives an answer to this question: 
{\sl The limit of number of particles $N$ going to infinity for energy functional (\ref{eq:H}) 
is the Thomas-Fermi model}. This means that
in this limit we recover again the leading order of a semiclassical approximation. So,
classical objects can be obtained when the number of particles in a system increases without bound
because {\sl the limit of a number of particles going to infinity coincides with the semiclassical
limit $\hbar\rightarrow 0$}.

This result can hold only if it is stable under perturbations. Indeed, the dynamical
equations for the Thomas-Fermi approximation are given by \cite{kirz}
\begin{eqnarray}
	\frac{\partial W}{\partial t}&+&  {\bf v}\cdot\nabla_x W-
    \frac{1}{m}\nabla_x (V + {\cal E}_F)\cdot\nabla_v W= 0 \\
	\Delta_2 V({\bf x},t) &=& 4\pi e^2 [\rho({\bf x},t)-Z\delta({\bf x})] \\
	\rho({\bf x},t) &=& \int W({\bf x},{\bf v},t)d^3v
\end{eqnarray}
being $W({\bf x},{\bf v},t)$ the Wigner function and 
${\cal E}_F=\frac{\hbar^2}{2m}(3\pi^2)^{\frac{2}{3}}\rho({\bf x},t)^{\frac{2}{3}}$ the Fermi energy.
This is just the classical limit of the Wigner-Poisson set of equations and the conclusion is
that classicality holds only if the initial preparation of the system permits the
effectiveness of Landau damping that dumps out any perturbation. Otherwise, higher order
quantum corrections come into play and the behavior is no more classical. This complete the
proof.

\section{Spin systems}

Spin systems are crucial for an understanding of the way a decoherence mechanism does work.
The reason for this relies on the fact that off-diagonal terms in the density matrix of
such systems appears to oscillate without any apparent decay. Indeed, recently this opened
up some discussion \cite{sch1,cas2}. The problem relies on the way numerical results should
be considered reliable when a large number of spins is interacting. The point is that a
sin function with a very large frequency, as happens when there are a large number of spins,
is very difficult to sample and becomes just a random number generator\cite{fra1}. Indeed the
physics is quite different as recently proved by experiments with organic molecular crystals
\cite{past1,past2,past3,ks}. They observe a Gaussian decay in time of quantum coherence that 
depends crucially on the number of spins of the system. So, we immediately see that numerical 
computations appear to be at odds with experiments. 

An explanation to the experimental results comes from a recently proved theorem due
to Hartmann, Mahler and Hess that can be stated in the following form
(here and in the following $\hbar=1$)

{\it If the many-body Hamiltonian $H=\sum_iH_i$ and the product state $|\phi\rangle$ satisfy
$\sigma_\phi^2\ge NC$
for all $N$ and a constant $C$ and if each $H_i$ is bounded as $\langle\chi|H_i|\chi\rangle\le C'$
for all normalized states $|\chi\rangle$ and some constant $C'$, then, for fidelity, the following holds
\begin{equation}
\nonumber
\lim_{N\rightarrow\infty}|\langle\phi|e^{-iHt}|\phi\rangle|^2=e^{-\sigma_\phi^2 t^2}.\nonumber
\end{equation}
}
and this is in perfect agreement with the experimental results as one has a Gaussian decay
with a decaying time inversely proportional to the number of interacting spins. This is an
intrinsic property of these systems.

Two points are relevant for this result. First of all there is a problem with recurrence. This
means that coherence can be recovered after some time. The above experiments did not check
this. We will see below for the Dicke spin-boson model how this fact is harmless for the appearance
of a classical world in the thermodynamic limit. The second question is linked to the
Gaussian evolution of the system. This result is well-known since the work of
Misra and Sudarshan about quantum time evolution \cite{ms1,ms2}. One has that at very short
times a quantum system display Gaussian time evolution. This means that in the thermodynamic
limit just the short time evolution is meaningful for spin systems. We will support also this
view with the discussion of the Dicke model.

\section{Dicke model}

A point that is generally neglected is that to realize a measurement on a quantum
system, the only means we have is electromagnetic interaction. The question that naturally
comes out in view of this simple consideration is if QED can manifest decoherence in the
thermodynamic limit in the low energy limit. We showed recently that in the low energy
limit several approximations hold for QED reducing the model to the the well-known
Dicke model \cite{fra2,fra3}. This model has Hamiltonian for a single radiation mode
\begin{equation}
H=\omega a^\dagger a+\frac{\Delta}{2}\sum_{i=1}^N\sigma_{3i}+g\sum_{i=1}^N\sigma_{1i}(a^\dagger+a)
\end{equation}
being $a,a^\dagger$ the annihilation and creation operators, $\omega$ the frequency of the mode,
$\Delta$ the separation between the levels of each two-level atom in the ensemble we are considering,
$g$ the coupling and $\sigma_{1i},\sigma_{3i}$ the Pauli spin matrices of the i-th atom.
We are interested on the effect of the limit $N\rightarrow\infty$ on the radiation mode.

One can show \cite{fra3} that when the limit $N\rightarrow\infty$ is taken, the Hamiltonian
ruling the dynamics of the model is
\begin{equation}
    H_F = \omega a^\dagger a + g\sum_{n=1}^N\sigma_{1i}(a^\dagger + a)
\end{equation}
that is, the model is integrable and the unitary evolution operator is
straightforwardly written down as
\begin{equation}
    U_F(t)=e^{-iH_Ft}=e^{i\hat\xi(t)}e^{-i\omega a^\dagger at}\exp[\hat\alpha(t)a^\dagger-\hat\alpha^*(t)a]
\end{equation}
being
\begin{equation}
    \hat\xi(t)=\frac{\left(\sum_{i=1}^N\sigma_{1i}\right)^2g^2}{\omega^2}(\omega t-\sin(\omega t))
\end{equation}
and
\begin{equation}
    \hat\alpha(t)=\frac{\left(\sum_{i=1}^N\sigma_{1i}\right)g}{\omega}(1-e^{i\omega t}).
\end{equation}
When the ensemble of two-level atoms is largely polarized (for the sake of simplicity
we take a fully polarization), for a generic radiation state $\sum_nc_n|n\rangle$ 
being $|n\rangle$ number states, one has
\begin{equation}
 \langle U_F(t)\rangle = e^{i\xi(t)}
	e^{-\frac{N^2g^2}{\omega^2}(1-\cos(\omega t))}
	\sum_{m,n}c_m^*c_ne^{-im\omega t}\frac{n!}{m!}
    \left[\frac{Ng}{\omega}(1-e^{i\omega t})\right]^{m-n}
	L^{m-n}_n\left[\frac{2N^2g^2}{\omega^2}(1-\cos(\omega t))\right]
\end{equation}
The real significant term in this solution is given by 
$e^{-\frac{N^2g^2}{\omega^2}(1-\cos(\omega t))}$ that for $N\rightarrow\infty$ goes
to zero except for very small times near the zeros of $1-\cos(\omega t)$ where it
reproduces a Gaussian. We have shown our main assertion that only short times are
meaningful for time evolution in the Dicke model in the thermodynamic limit \cite{fra1}.
This describes a collapsing Gaussian evolution in agreement to Misra and Sudarshan
analysis.

As the zeros of the function $1-\cos(\omega t)$ are recurring with a period $2\pi/\omega$
we face here the problem of recurrence. The crucial point for this case is the width of the
recurring Gaussian. This is given by a time $1/Ng$ that becomes decreasingly small
with $N$ becoming increasingly large reaching unphysical values for a macroscopic
ensemble of atoms. This warrants that no recurrence is ever observed and a truly classical
state is reached and maintained as also happens for spin systems. Numerical evidence
exists for these results \cite{eb1,eb2}.

Finally, we point out that the opposite limit of a large number of photons has been
discussed by Gea-Banacloche \cite{gb1,gb2} and experimentally proved by Haroche's group
\cite{har1,har2}.

\section{Conclusions}

We have presented a lot of theoretical, mostly theorems, and some experimental evidence
supporting our view that the thermodynamic limit has a deep meaning also in quantum
mechanics. A lot of experimental activity could be accomplished to understand how relevant
is such an approach to real systems. E.g. Haroche's group tested QED for a large number of
photons but nobody has done the same for a large number of atoms so far.
On the other side, the situation of spin systems appears more promising in the short term
as also happens in interference of large molecules \cite{zeil}. In this latter cases
possible interesting news can be expected. The success of this view will prove more
striking than ever the link between such different approaches as quantum and statistical
mechanics.

\ack 
I would like to thank the organizers and specially Prof. Elze for giving me the
opportunity to contribute to this beautiful conference.
\newline


\begin{thebibliography}{99}
\bibitem{zeh} Giulini D, Joos E, Kiefer C, Kupsch J, Stamatescu I-O and Zeh H D 1996
{\it Decoherence and the Appearance of a Classical World in Quantum Theory},
(Berlin:Springer).
\bibitem{hepp} Hepp K 1972 {\it Helv. Phys. Acta} {\bf 45}, 237.
\bibitem{boni} Bonifacio R, Olivares S, Tombesi P and Vitali D 2000 
{\it Phys. Rev. A} {\bf 61}, 053802.
\bibitem{cast} Castagnino M and Lombardi O 2005 {\it Phys. Rev. A} {\bf 72}, 012102.
\bibitem{ghir} Bassi A and Ghirardi G C 2003 {\it Phys. Rept.} {\bf 379}, 257.
\bibitem{zure} Zurek W H 1998 {\it Phys. Scri.} {\bf T76}, 186. 
\bibitem{ball} Wiebe N and Ballentine L E 2005 {\it Phys. Rev. A} {\bf 72}, 022109.
\bibitem{schl} Schlosshauer M 2006 {\it Preprint } quant-ph/0605249.
\bibitem{fra1} Frasca M 2006 {\it Mod. Phys. Lett. B} {\bf 20}, 1059.
\bibitem{lau} Laughlin R B and Pines D 2000 {\it Proc. Natl. Acad. Sci.} {\bf 97}, 28.
\bibitem{nt} Ring P and Schuck P 2004 {\sl The Nuclear Many-Body Problem},  
(Berlin:Springer).
\bibitem{str} Dalfovo F, Giorgini S, Pitaevskii L P and Stringari S 1999 {\it Rev. Mod. Phys.} {\bf 71}, 463.
\bibitem{par} Parr R G and Yang Weitao 1995 {\sl Density-functional Theory of Atoms and Molecules},
(Oxford:Oxford University Press).
\bibitem{sch} Kleinert H 2004 {\it Path Integrals in 
Quantum Mechanics, Statistics, and Polymer Physics, and Financial Markets}, (Singapore: World Scientific).
\bibitem{ls1} Lieb E H and Simon B 1973 Phys. Rev. Lett. {\bf 31}, 681.
\bibitem{ls2} Lieb E H and Simon B 1977 Adv. in Math. {\bf 23}, 22. 
\bibitem{kirz} Kirzhnitz D A, Lozovik Yu E and Shpatakovskaya G V 1975 {\it Sov. Phys. Usp.} {\bf 18}, 649.
See also \cite{nt}.
\bibitem{sch1} Schlosshauer M 2005 {\it Phys. Rev. A} {\bf 72}, 012109.
\bibitem{cas2} Castagnino M and Laura R 2005 {\it Preprint} quant-ph/0512131.
\bibitem{past1} Levstein P R, Usaj G and Pastawski H M 1998 {\it Preprint} cond-mat/9803047.
\bibitem{past2} Levstein P R, Usaj G and Pastawski H M 1998 {\it Mol. Phys.} {\bf 95}, 1229. 
\bibitem{past3} Levstein P R, Usaj G and Pastawski H M 1998 {\it J. Chem . Phys.}  {\bf 108}, 2718.
\bibitem{ks} Krojanski H G and Suter D 2004 {\it Phys. Rev. Lett.} {\bf 93}, 090501.
\bibitem{hmh} Hartmann M, Mahler G and Hess O 2004 {\it Lett. Math. Physics} {\bf 68}, 103 
({\it Preprint} math-ph/0312045).
\bibitem{ms1} Misra B and Sudarshan E C G 1977 J. Math. Phys. {\bf 18}, 756.
\bibitem{ms2} Chiu C B, Sudarshan E C G and Misra B 1977 Phys. Rev. D {\bf 16}, 520.
\bibitem{fra2} Frasca M 2003 Annals Phys. {\bf 306}, 193.
\bibitem{fra3} Frasca M 2004 Annals Phys. {\bf 313}, 26.
\bibitem{eb1} Emary C and Brandes T 2003 Phys. Rev. Lett. {\bf 90}, 044101.
\bibitem{eb2} Emary C and Brandes T 2003 Phys. Rev. E {\bf 67}, 066203.
\bibitem{gb1} Gea-Banacloche J 1990 Phys. Rev. Lett. {\bf 65}, 3385.
\bibitem{gb2} Gea-Banacloche J 1991 Phys. Rev. A {\bf 44}, 5913.
\bibitem{har1} Bertet P, Osnaghi S, Rauschenbeutel A, Nogues G, Auffeves A, Brune M,
Raimond J M and Haroche S 2001 Nature {\bf 411}, 166.
\bibitem{har2} Auffeves A, Maioli P, Meunier T, Gleyzes S, Nogues G, Brune M,
Raimond J M and Haroche S 2003 Phys. Rev. Lett. {\bf 91}, 230405.
\bibitem{zeil} Hackerm\"uller L, Uttenthaler S, Hornberger K, Reiger E, Brezger B, Zeilinger A and Arndt M 
2003 Phys. Rev. Lett. {\bf 91}, 090408.
\end{thebibliography}
\end{document}